\documentclass[aps,pra,preprint,groupedaddress]{revtex4-1}
\usepackage{amsmath,amsfonts,amsthm} 
\usepackage{graphicx}
\usepackage{xcolor}
\usepackage{physics}
\usepackage{hyperref}

\begin{document}

\title{Spatial structure of the pair wavefunction and the density correlation functions at the BEC-BCS crossover}

\author{J.C. Obeso-Jureidini and V. Romero-Roch\'in}
\affiliation{Instituto de F\'isica, Universidad Nacional Aut\'onoma de M\'exico \\
Apartado Postal 20-364, 01000 Cd. M\'exico, Mexico}

\date{\today}

\begin{abstract}

By an exact numerical calculation of the BCS pair wavefunction and the density correlation functions both between atoms in the same and in different spin states, we extract the spatial large distance behavior of the respective functions. After different initial transients, those distributions show an algebraic dependence accompanied with their own exponential decay and  a well defined periodic oscillatory behavior. While in general, in the BCS side there are long-range correlations and in the BEC region the behavior is dominated by tight pairs formation, each distribution shows its own overall behavior.
We derive analytic expressions for the mean pair size and the correlation lengths of the same and different density correlation functions. The whole analysis yields a quite complete description of the spatial structure of the superfluid along the crossover.

\end{abstract}

\maketitle

\section{Introduction}

The physics of a mixture of two fermionic species with tunable interactions remains as a subject of current interest. Such a model system is ubiquitous in a wide range of physical phenomena, ranging from its realization in ultracold gases, commonly found now in many laboratories,\cite{Mukherjee,Zwierlein,Regal, Kinast, Bourdel, Chin-Science,Hulet,Zwierlein2,Esslinger,Roati1} as one explanation of high-temperature superconductivity \cite{Randeria,Micnas,Drechsler,Haussmann,Pistolesi}  and to a description of nuclear matter in certain stars \cite{Strinati-review,Baker,Lombardo,Andrenacci,Jin}. The overall physical picture is of a gas of large, overlapping Cooper pairs of atoms in the Bardeen-Cooper-Schriefer (BCS) regime and, in the other extreme, a Bose-Einstein condensate (BEC) of tight molecules, with a smooth crossover occurring in the neighborhood within the two-body collision scattering resonance and the sign change of the chemical potential \cite{Blatt, Eagles, Leggett, Nozieres}. The theoretical work on this subject is already overwhelming for instance see Refs. \cite{Strinati-review,Ketterle-review,Giorgini-Stringari} as excellent reviews. An important and apparently simple question concerns the size of the pairs of atoms of the two different species along the crossover and, although there are already several discussions to answer this question,\cite{Pistolesi,Ortiz,Bertsch,Ketterle-review,Marini} it does not and perhaps cannot have a simple or sole answer since this is clearly a many-body problem. Here, we readdress this question and pose it more as an analysis of the spatial structure of the mixture rather than just a question concerning the size of the pairs. For this goal, within the mean-field BCS-Leggett theory, by an accurate calculation of their spatial dependence, we study the following two-body distributions: the BCS pair wavefunction, the density correlation function of atoms of different species and the density correlation function of like particles. The three of them are obtained directly from the standard variational solution of the problem at zero temperature \cite{Leggett}. In addition, we introduce a postulated ``pair-binding'' wavefunction or distribution, based on the pair-binding energy \cite{Bethe-Peierls}, as a reference for the previous three functions.  

Our analysis is  based mainly on a novel  numerical evaluation of the spatial dependence for the mentioned distributions, which allows us to accurately fit their exponential decay length as well as their oscillatory wavenumbers and phases for long distances. In addition, we provide exact analytical expressions for the correlation lengths of all the distributions, being defined as the normalized second moment of each distribution \cite{Strinati-review}. These characteristic lengths and the associated oscillatory behavior yield a very complete picture of the physical nature of the pairing phenomenon along the whole crossover.  
We first observe that the three distribution functions behave differently in some regions while similarly in others. For instance, in the BCS side both correlation functions, of the same and different species, show the expected long-range behavior with divergent coherence or correlation lengths, while the BCS probability distribution shows a finite average pair radius \cite{Ortiz}. The three previous results put together indicate that, in the BCS limit, although the pair mean size is finite there are pairs of all sizes as all distributions decay algebraically. 
On the other limit, at BEC, now the unlike particle correlation function and the BCS pair probability distribution behave almost equally, while the same species correlation function decays extremely fast. This is certainly another indication of the formation of a gas of bosonic molecules where there are no Pauli-blocking  correlations of atoms in the same spin state. The binding-pair distribution is a reference that fits quite well the envelope of the calculated distributions in the crossover. Regarding the oscillatory behavior, we find that in the long distance limit the three distributions spatially oscillate with the same wavenumber that tends to the Fermi momentum in the BCS limit and vanishes in the BEC one. The like and unlike atomic correlations oscillate perfectly out of phase, showing a nested structure that becomes more evident in the BCS limit. While such a structure remains in the BEC side, it is evidently arrested by the fast exponential decay and low frequency oscillations. The BCS wavefunction oscillates with the same frequency but its relative phase changes along the crossover.

The article is organized as follows. First we briefly review the Leggett-BCS mean-field model as a reference for the
calculation and discussion of the pair BCS wavefunction and the correlation functions in the following sections. In
section III we discuss the mentioned distributions and discuss how, by deforming the Fourier transforms contours in the $k$-complex plane, we are able to calculate numerically the distributions for any pair spatial separation. In section IV we analyze the average pair radius and the correlation lengths by means of exact analytical expressions. We conclude with some final remarks. In the Appendix we give the 
essential details for the contour deformation of section III and the exact analytical expressions that we
provide along in the text.

\section{BEC-BCS mean-field contact interaction mixture gas}

We consider the usual contact interaction many-body Hamiltonian of a balanced gas mixture of fermion atoms in two hyperfine states $\sigma = \uparrow, \downarrow$, using the grand potential $\hat{\Omega}= \hat{H}-\mu \hat{N}$, 
\begin{equation}\label{eq_expectacion1_h-mu}
\hat{\Omega} =  \sum_{\vec{k}, \, \sigma} ( \varepsilon_{\vec{k}} - \mu ) c^\dagger_{\vec{k}\sigma} c _{\vec{k} \sigma} + \frac{g}{V} \sum_{\vec{k}_1} \sum_{\vec{k}_4} \; c^\dagger _{\vec{k}_1 \uparrow}c^\dagger _{- \vec{k}_1\downarrow} c _{-\vec{k}_4 \downarrow} c _{\vec{k}_4 \uparrow} ,
\end{equation}
where $g = 4 \pi \hbar^2 a /m$ is the interaction constant, $a$ the $s$-wave scattering length, $\varepsilon_{\vec{k}}= \hbar^2 k^2/2m$, $V$ is the volume of the sample and the sums are over all wave vectors. To find the ground state we use the BCS-Leggett variational method \cite{BCS-superconductivity,Leggett,Eagles}, although one can also rely on the mean field Method \cite{Strinati-review}, or the exact solution in the thermodynamic limit \cite{Ortiz}. 
The BCS-Leggett variational approach introduces the BCS wave function to minimize the Grand Potential
\begin{equation}\label{eq_estado_bcs_introducido}
\ket{\Psi_{BCS}} = \prod_{\vec{k}} \big(u_{\vec{k}} + v_{\vec{k}} c^{\dagger}_{\vec{k}\uparrow} c^{\dagger}_{-\vec{k}\downarrow}  \big) \ket{0},
\end{equation}
with $\hat c^\dagger_{\vec k\sigma}$ creation operators of fermionic atoms with momentum $\vec k$ and spin $\sigma = \uparrow, \downarrow$. The variational parameters satisfy the normalization condition $|u_{\vec{k}}|^2 + |v_{\vec{k}}|^2=1$ and are given by
\begin{equation}
\left\{ \begin{array}{c}
u_k^2 \\ v_k^2
\end{array}\right\} = \frac{1}{2} \left[1 \pm \frac{\epsilon_{\vec k}-\mu}{\sqrt{(\epsilon_{\vec k}-\mu)^2 + \Delta^2}} \right]\label{uvk}
\end{equation}
where the gap $\Delta$ has been introduced. For completeness and for further purposes below, we first give analytic expressions for the thermodynamic variables: the regularized gap $\Delta$, the number  $N = \langle \hat N\rangle$ and the gas ground state energy $E_0 = \langle \hat H \rangle$ equations. As it is shown in Appendix, these quantities and the characteristic lengths defined below, can all be expressed in terms of hypergeometric functions ${\rm F}= {\rm _2F_1}(a,b;c,(1-z)/2)$, which in turn in some cases can also be expressed in terms of Legendre functions ${\rm P}_\mu^\nu(z)$, with $z$ a natural dimensionless quantity,
\begin{equation}
z = - \frac{\mu}{\sqrt{\mu^2 + \Delta^2}} .
\end{equation}
We note that $-1 < z <+1$, and that the deep BCS limit, $a \to 0^-$, is $z \to -1$, while the BEC one, $a \to 0^+$, is $z \to +1$. In all expressions we use the thermodynamic limit $\sum_{\vec k} \to \frac{V}{(2\pi)^3}\int d^3k$. The so-called regularized gap equation expresses the scattering length $a$ in terms of the the chemical potential $\mu$ and the gap $\Delta$,
\begin{eqnarray}\label{eq_ecuacion_gap_k}
-\frac{m}{4 \pi \hbar^2 a} &=&  \frac{1}{2V}  \sum_{\vec k} \Bigg(\frac{1}{\sqrt{(\varepsilon_{\vec{k}}- \mu)^2 + \Delta^2}} - \frac{1}{\varepsilon_{\vec{k}}} \Bigg) \nonumber \\
& = & -\frac{1}{8\pi} \left(\frac{2m}{\hbar^2}\right)^{3/2} \left(\mu^2 + \Delta^2\right)^{1/4} {\rm P}_{\frac{1}{2}}(z). \label{gap}
\end{eqnarray}

The number, or rather the particle density equation, $n = N/V$,
\begin{eqnarray}\label{eq_ecuacion_num_k}
n &=&  \frac{1}{V}  \sum_{\vec k} \Bigg( 1 - \frac{\varepsilon_{\vec{k}}- \mu }{\sqrt{(\varepsilon_{\vec{k}}- \mu)^2 + \Delta^2 }} \Bigg) \nonumber  \\
& = & \frac{1}{4\pi} \left(\frac{2m}{\hbar^2}\right)^{3/2} \left(\mu^2 + \Delta^2\right)^{3/4} \left[- {\rm P}_{\frac{3}{2}}(z)+z {\rm P}_{\frac{1}{2}}(z) +\right] \label{density}
\end{eqnarray}
Since the gas mixture is balanced, the number of $\uparrow$ atoms equal those of spin  $\downarrow$, being $N/2$.

In the same way, the total energy density $e_0 = E_0/V$ is,
\begin{eqnarray}\label{eq_ecuacion_ene_k}
e_0 &=&  \frac{1}{V}  \sum_{\vec k} \Bigg(\varepsilon_{\vec{k}} - \frac{\varepsilon_{\vec{k}}(\varepsilon_{\vec{k}}-\mu) - \Delta^2/2}{\sqrt{(\varepsilon_{\vec{k}}- \mu)^2 + \Delta^2}} \Bigg). \nonumber \\
& = &\frac{1}{4\pi} \left(\frac{2m}{\hbar^2}\right)^{3/2} \left(\mu^2 + \Delta^2\right)^{5/4}
\left[{\rm P}_{\frac{5}{2}}(z) - z {\rm P}_{\frac{3}{2}}(z) + \frac{1}{2}(1 - z^2) {\rm P}_{\frac{1}{2}}(z)\right]\label{energy} .
\end{eqnarray}
The above quantities have already been found in similar forms in other references \cite{Ortiz,Marini}. We do include them here for the analysis performed below. 

Combination of the first two equations, (\ref{gap}) and (\ref{density}), into the third one, (\ref{energy}), and using a recursion relation of Legendre functions,
yields the following closed expression of $e_0$ in terms of the four thermodynamic quantities $n$, $\mu$, $a$ and $\Delta$, 
\begin{equation}
e_0 = -\frac{1}{5} \left[\frac{m}{4 \pi \hbar^2 a} \Delta^2 - 3 \mu n\right].
\end{equation}
As it will be of use below, we can identify the  binding energy {\it per} pair, as \cite{Ortiz}
\begin{eqnarray}
\epsilon_b & = & \frac{2}{N} \left(E_F - E_0\right) \nonumber \\
& = & \frac{2}{5n} \left[3 (\epsilon_F - \mu)n + \frac{m}{4 \pi \hbar^2 a} \Delta^2 \right], \label{eb}
\end{eqnarray}
where $E_F = (3/5) N \epsilon_F$ is the ground state energy of $N$ free fermions, with $\epsilon_F = \hbar^2 k_F^2/2m$ and $k_F = \left(3 \pi^2 n\right)^{1/3}$ the Fermi energy and momentum of the free gas. In the mean field method the quasiparticle spectrum suggests that the energy to create such an excitation with ${k} \approx {0}$ is given by
\begin{equation}
\epsilon_{spec} = \sqrt{\mu^2+ \Delta^2} - \mu. \label{espec}
\end{equation}
This quantity has been used to analyze RF-spectroscopy of pair dissociation \cite{Ketterle-review,tinkham}. As we see below, this energy is very close to $\epsilon_b$ along the whole crossover.

As it is discussed in many texts, Eqs. ({\ref{gap}) and (\ref{density}) can be used to solve $\Delta$ and $\mu$ as functions of $n$ and $a$, thus allowing to express all the physical quantities of the crossover in terms of the latter as the independent thermodynamic variables. Further, it is common and useful to write the above expressions in dimensionless forms. For this, we introduce the notation $\tilde \epsilon = \epsilon/\epsilon_F$ for energies and $\tilde l = k_F l$ for lengths. In Figure \ref{Fig1} we plot $\tilde \mu$, $\tilde \Delta$, $\tilde \epsilon_b$ and $\tilde \epsilon_{spec}$ as functions of $1/k_F a$, obtained from Eqs. (\ref{gap}), (\ref{density}), (\ref{eb}) and (\ref{espec}), as the reference variables for the analysis below. Also, as obtained from the previous equations, see also Ref. \cite{Ortiz}, we quote here the asymptotic values of $\tilde \Delta$, $\tilde \mu$ and $\tilde \epsilon_b$ in the BCS and BEC limits, $1/k_Fa \to -\infty$ and $1/k_Fa \to +\infty$ respectively,
\begin{eqnarray}
&\tilde \Delta \approx 8 e^{-2} e^{\frac{\pi}{2 k_F a}}; \>\>\>\> \tilde \mu \approx 1 + 8 e^{-4} \left[\frac{\pi}{k_F a}-5\right] e^{\frac{\pi}{k_F a}}; \>\>\>\> \tilde \epsilon_b \approx 48 e^{-4} e^{\frac{\pi}{k_F a}}; \>\>\>\> {\rm for}\>\>1/k_Fa \to -\infty& \nonumber \\
&\tilde \Delta \approx \left(\frac{16}{3 \pi k_Fa}\right)^{1/2}; \>\>\>\> \tilde \mu \approx  - \frac{1}{(k_Fa)^2}; \>\>\>\> \tilde \epsilon_b \approx \frac{2}{(k_Fa)^2};  \>\>\>\> {\rm for}\>\>1/k_Fa \to +\infty&\label{limits}
\end{eqnarray}

From the above exact expressions, Eqs.(\ref{eq_ecuacion_gap_k}) and (\ref{eq_ecuacion_num_k}), one can find the known special values, first, at unitarity $1/k_Fa =0$, $\tilde \mu \approx 0.59061$ and $\tilde \Delta \approx 0.68640$; and, second, at $\tilde \mu = 0$, $1/k_Fa \approx 0.55315$ and $\tilde \Delta \approx 1.0518$. 

\begin{figure}[htbp]
\begin{center}
\includegraphics[width=0.5\linewidth]{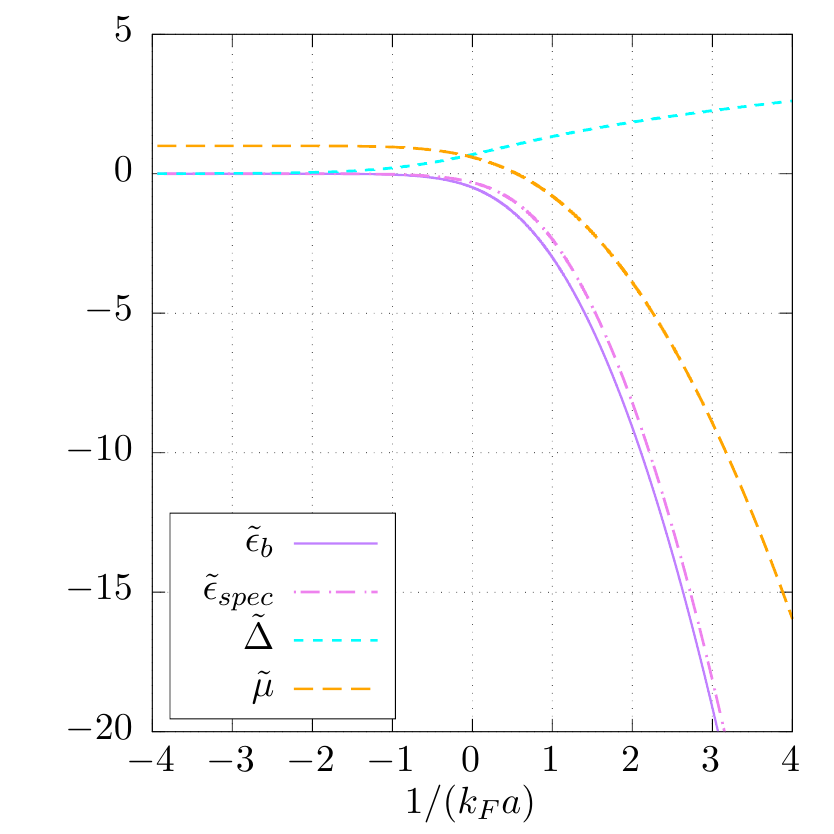}
\end{center}
\caption{(Color online) Dimensionless chemical potential $\tilde \mu = \mu/\epsilon_F$, gap $\tilde \Delta = \Delta/\epsilon_F$, binding pair energy $\tilde \epsilon_b = \epsilon_b/\epsilon_F$ and spectroscopic threshold $\tilde\epsilon_{spec}= \epsilon_{spec}/\epsilon_F$, as functions of $1/k_F a$.} \label{Fig1}
\end{figure}

\section{BCS-pair wavefunction and density correlation functions}

As described in the Introduction, the spatial structure of the many-body BCS state cannot be solely pinned on a single quantity, but rather, one can look at several relevant quantities that, put together, yield a more complete picture without the need of compromising the concept of what a Cooper pair really is. For this, we look at the three 
two-body quantities that can be extracted from the variational parameters $u_{\vec k}$ and $v_{\vec k}$. These are, the (a) BCS-pair wavefunction $\phi_{BCS}(\vec r)$; (b) the density correlation between different spin species $G_{\uparrow \downarrow}(\vec r)$; and, the density correlation between same species $G_{\uparrow \uparrow}(\vec r)$. The latter being equal to the $\downarrow \downarrow$ correlation due to the balance of  the mixture.  
Although these quantities have been fully or partially addressed in the literature, 
the improvement here presented resides on the fact that we are able to accurately calculate their spatial dependence for any distance. In particular, we study the long-range behavior of all of them, yielding not only their {\it exponential} decaying lengths $\chi$ but also their respective main oscillatory wavelength or wavenumber $\kappa$ as well as their relative phases. In addition, in the following section, we also calculate analytically the mean pair radius and the correlation lengths defined as the second moment of the corresponding distributions. 

As it was already pointed out by Leggett in his seminal work, \cite{Leggett} the BCS state, spatially, is the antisymmetric superposition of $\uparrow \downarrow$ pairs, with every pair being in a two-body state given by the following {\it unnormalized} wavefunction
\begin{equation}
\phi_{\rm BCS}(\vec{r}) = \frac{1}{(2\pi)^3} \int d^3 \vec{k} \> e^{i \vec{k} \cdot \vec{r}}\> \frac{v_{\vec{k}}}{u_{\vec{k}}} .\label{phi}
\end{equation}
As it is our claim here, however, the many-body spatial structure also depends on the next in importance density correlation functions of like and unlike pairs.} To calculate the latter, we recall that 
the density operator, at $\vec r$ of spin $\sigma = \uparrow$ or $\downarrow$, is given by $\hat n_{\sigma}(\vec r) = \hat \psi^\dagger_\sigma(\vec r) \hat \psi_\sigma(\vec r)$, where the particle annihilation operator is
\begin{equation}
\hat \psi_\sigma(\vec r) = \frac{1}{\sqrt{V}} \sum_{\vec k} e^{i\vec k \cdot \vec r} \> \hat c_{\vec k, \sigma}.
\end{equation}
Then, the density correlation function of antiparallel spins  $\uparrow \downarrow$ is found to be
\begin{equation}\label{eq_correlador_antiparalelo}
\begin{split}
G_{\uparrow \downarrow}(\vec{r},\vec r^{\> \prime}) = \langle \hat n_\uparrow(\vec{r}) \hat n_\downarrow(\vec r^{\> \prime}) \rangle -\langle \hat n_\uparrow(\vec{r}) \rangle \langle  \hat n_\downarrow(\vec r^{\> \prime}) \rangle\\
=  \left|g_{\uparrow \downarrow}(\vec{r}-\vec r^{\> \prime})\right|^2,
\end{split}
\end{equation}
with $g_{\uparrow \downarrow}(\vec{r})$ the Fourier transform of $v_{\vec{k}} u_{\vec{k}}$,
\begin{equation}
g_{\uparrow \downarrow}(\vec{r}) =  \frac{1}{(2\pi)^3} \int d^3 \vec{k} \> e^{i \vec k \cdot \vec r} v_{\vec{k}} u_{\vec{k}} .\label{gpm}
\end{equation}
As we will discuss further below, this quantity has also been identified as the ``pair wavefunction'' in several studies, see Refs. \cite{Ketterle-review,Strinati-review}. Here, for definiteness we shall keep it associated to the ${\uparrow \downarrow}$ density correlations.

And, thirdly, the density correlation with parallel $\uparrow \uparrow$ spins (equal to $\downarrow \downarrow$) can be found to be,
\begin{equation}\label{eq_correlador_paralelo}
\begin{split}
G_{\uparrow \uparrow}(\vec{r},\vec r^{\> \prime}) = \langle  \hat{n}_{\uparrow}(\vec{r}) \hat{n}_{\uparrow}(\vec r^{\> \prime}) \rangle \,-\, \langle  \hat{n}_{\uparrow}(\vec{r}) \rangle \langle \hat{n}_{\uparrow}(\vec r^{\> \prime}) \rangle\\
= \frac{n}{2} \delta^3(\vec{r}-\vec r^{\> \prime}) - \left|g_{\uparrow \uparrow}(\vec{r}-\vec r^{\> \prime})\right|^2,
\end{split}
\end{equation}
where now $g_{\uparrow \uparrow}(\vec{r})$ is the Fourier transform of $v^2_{\vec{k}}$,
\begin{equation}
g_{\uparrow \uparrow}(\vec{r}) = \frac{1}{(2\pi)^3} \int d^3 \vec{k} \> e^{i \vec k \cdot \vec r} v^2_{\vec{k}} .\label{gpp}
\end{equation}

The density correlation with parallel spins considered here differs from the one commonly used in the literature \cite{Giorgini-Stringari,Pessoa-contact}, the present one including the same site contribution of the spin. This is important to note because the behavior is completely different from those previously used. Hence, at the two-body level, the structure of BCS state is contained in the quantities $v_{\vec k}/u_{\vec k}$, $v^2_{\vec k}$ and $u_{\vec k}v_{\vec k}$. It is of interest to point out that the negative, or anticorrelation  sign in the second term of the equal spin correlation function, reflects the Pauli exclusion principle just as in the ideal Fermi gases \cite{LandauII}.

As we can see from Eqs. (\ref{phi}), (\ref{gpm}) and (\ref{gpp}), the knowledge of $\phi_{BCS}(\vec r)$, $g_{\uparrow \uparrow}(\vec{r})$ and $g_{\uparrow \downarrow}(\vec{r})$ requires to perform a Fourier transform. These three expressions can be generically written as
\begin{equation}
f(\vec r) =  \int d^3 k \> e^{i \vec k \cdot \vec r} \>{\cal F}(\vec k), \label{ftilde}
\end{equation}
with ${\cal F}(\vec k)$ being $v_{\vec k}/u_{\vec k}$, $v^2_{\vec k}$ and $v_{\vec k} u_{\vec k}$, respectively. Since $u_{\vec k}$ and $v_{\vec k}$ depend on $k = |{\vec k}|$ only, see Eq.(\ref{uvk}), the functions $f(\vec r)$ depends on $r = |\vec r|$ only and can be generally written as
\begin{equation}
f(r) = \frac{1}{4 \pi^2 i r} \int_{-\infty}^{\infty} e^{ikr} \> k {\cal F}(k) dk .\label{fr}
\end{equation}
Thus, the task is reduced to perform the one-dimensional Fourier transform of $k {\cal F}(k)$. Since ${\cal F}(k)$ is a real even function of $k$, the functions $f(r)$ are also real. Calculating these Fourier transforms may suffice the use of a common fast-Fourier transform. However, since the functions ${\cal F}(k)$ decay algebraically for large $k$, the numerical precision of $f(r)$ is severely limited making it very hard to find its large $r$ behavior. Here we present an alternative approach that allows for an accurate calculation of those Fourier transform for any value of $r$, short and very large. This gives rise to precise fits of the exponential decay length and for the oscillatory component in the long distance regime. The mathematical details are given in Appendix and here we just present the main aspects of the corresponding calculations.

\begin{figure}[htbp]
\begin{center}
\includegraphics[width=1.0\linewidth]{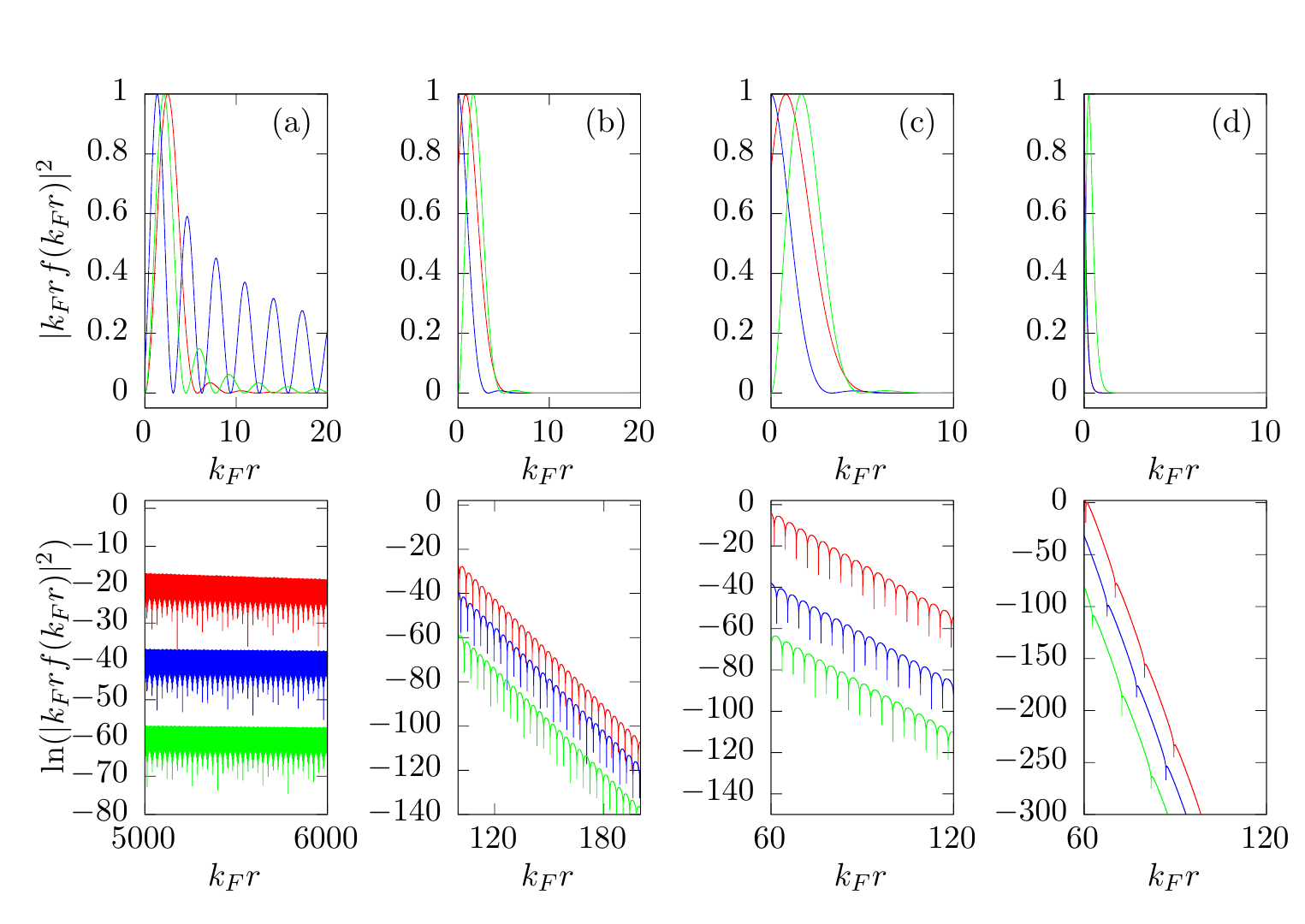}
\end{center}
\caption{(Color online) Distributions $|r f(r)|^2$ as functions of $k_F r$, for chosen values of $1/k_F a$ before and after the resonance, (a) $1/k_Fa = -5.0$; (b) $1/k_Fa = -0.0046$; (c) $1/k_Fa = 0.0064$; (d) $1/k_Fa = 4.0$. The lower panel is in semi-logarithmic scale. Upper (red) lines correspond to $|r \phi_{\rm BCS}(r)|^2$; middle (blue) lines $|r g_{\uparrow \downarrow}(r)|^2$; and lower (green) lines $|r g_{\uparrow \uparrow}(r)|^2$.} \label{fig2}
\end{figure}

As it is usual when dealing with this model, all analytical and numerical complications arise from the square root $\sqrt{(\epsilon_{\vec k}-\mu)^2 + \Delta^2}$ term in the expressions for $u_{\vec k}$ and $v_{\vec k}$, see Eq.(\ref{uvk}). In the $k$-complex plane this square root gives rise to four branch points with their corresponding branch cuts. Hence, the integral $f(r)$ in Eq.(\ref{fr}) can be extended to the $k-$complex plane, followed by a deformation of the countour integral, yielding an alternative expression for $f(r)$ that can be accurately numerically integrated for large values of $r$. As an example, we write here the expression for the pair wavefunction $\phi_{\rm BCS}(r)$ only,
\begin{eqnarray}
 \phi_{\rm BCS}(r) = -\frac{k^3_F}{\pi^2 \tilde \Delta \> k_F r} && \int_{\tau_0}^\infty \Bigg( \frac{2\tau^2 - \tilde \mu}{\sqrt{\tau^2 - \tilde \mu}} \Bigg) \sqrt{  (2 \tau^2 -\tilde \mu)^2 - (\tilde \Delta^2 + \tilde \mu^2)} \times \nonumber \\ 
 && \exp \Big[- \sqrt{\tau^2 - \tilde \mu} \; k_F r \Big] \>\sin ( k_F r \>\tau  )  d\tau
\end{eqnarray}
where $\tau_0 = ((\tilde \mu + (\tilde  \mu^2 +\tilde \Delta^2)^{1/2})/2)^{1/2}$. While the above expression may look complicated, it actually converges very fast and accurately due to the exponentially decaying term in the integrand. The expressions for $g_{\uparrow \uparrow}(\vec{r})$ and $g_{\uparrow \downarrow}(\vec{r})$ also contain the same exponential term with an analogous fast convergence. Although one can calculate any of the above distributions for any value of $r$, as shown in Fig. \ref{fig2}, we mainly concentrate on their long-distance behavior as we now discuss. In the literature this three functions have been studied for short distances only \cite{Giorgini-Stringari,Pessoa-contact}.

{\bf Exponential decaying lengths}. The upper panel of Figure \ref{fig2} shows the functions $|r\phi_{\rm BCS}(\vec r)|^2$, $|rg_{\uparrow \uparrow}(\vec{r})|^2$ and $|r g_{\uparrow \downarrow}(\vec{r})|^2$ for small values of $r$, where differences are evident. However, once certain transient has been passed, their asymptotic shape, for large values of $r$ appear quite similar, as can be seen in the corresponding lower panel of Figure \ref{fig2}. The 
latter are plots in semi-log scale showing an oscillatory behavior with characteristic wavenumbers $\kappa$ and phases $\varphi$, that will be detailed below. On top of the oscillations there is a clear exponential decay, with a characteristic {\it exponential decay} length $\chi$, that certainly depends on $k_F a$. Accurate fits of $\chi$, $\kappa$ and $\varphi$ for each distribution, yield a generic function for the the three distributions of the form,
\begin{equation}
f^2(r) \approx \frac{\rm Const}{r^2} e^{-\frac{\sqrt{2}\> r}{\chi}} {\cal P}(\kappa r + \varphi) \>\>\>\>r \gg k_{F}^{-1}
\end{equation}
with ${\cal P}(\kappa r + \varphi)$ a periodic function of $r$, with wavelength $2\pi/\kappa$ and phase $\varphi$. We would like to point out that this full functional form has not been previously studied in all the crossover region for the three functions that we are considering.

The previous fit suggests to associate an additional function that helps describing the spatial envelope of the pairs. This we call the {\it binding pair} function, which we write as
\begin{equation}
|\Phi_b(r)|^2 = \frac{\rm Const}{r^2} e^{-\sqrt{2}\> r/\chi_b} ,
\end{equation}
valid for all values of $r$, with the exponential decay length $\chi_b$ given by 
\begin{equation}
\chi_{b} \equiv \left(\frac{\hbar^2}{2 m \epsilon_b}\right)^{1/2},\label{chib}
\end{equation}
where $\varepsilon_b$ is the binding energy shown in equation (\ref{eb}). 
A {\it spectroscopic} length $\chi_{spec}$ \cite{Ketterle-review,schunckdetermination} associated to the threshold energy $\epsilon_{spec}$, as given in Eq. (\ref{espec}), can also be used for comparison. In fact, we have found $\chi_{\uparrow \downarrow}$ and $\chi_{\uparrow \uparrow}$ to be well fitted by $\chi_{spec}$. However, we do not consider it here explicitly since this length is very close to $\chi_b$ along the whole crossover, as can be inferred from Fig. \ref{Fig1}.

\begin{figure}[htbp]
\begin{center}
\includegraphics[width=0.8\linewidth]{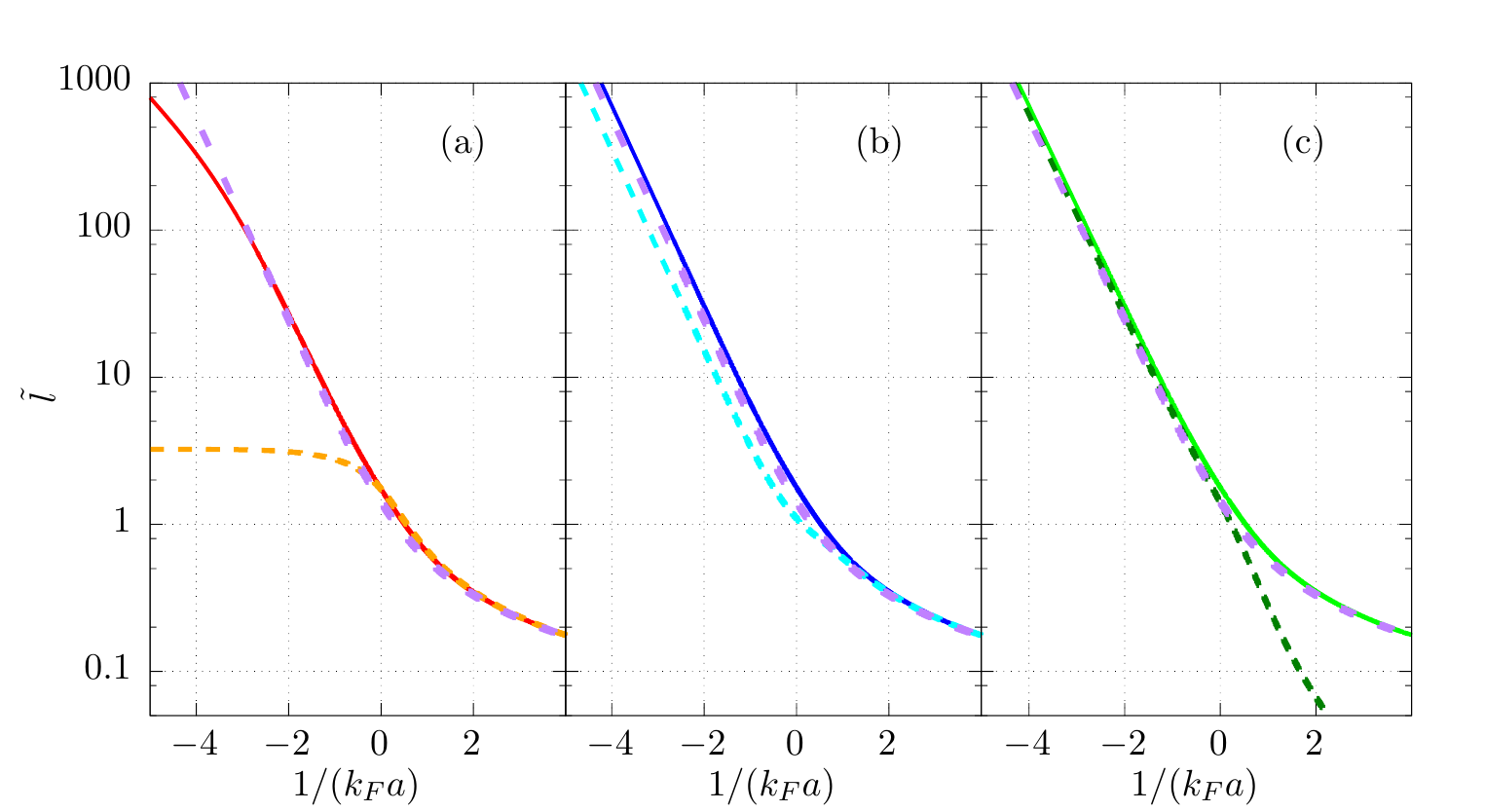}
\end{center}
\caption{(Color online) Characteristic lengths $\tilde l = k_F l$ as functions of $1/k_Fa$, for the three distributions (a) $|\phi_{BCS}(r)|^2$; (b) $|g_{\uparrow \downarrow}(r)|^2$; and (c) $|g_{\uparrow \uparrow}(r)|^2$. Solid lines correspond to the exponential decay lengths (a) $\tilde \chi_{BCS}$, (b) $\tilde \chi_{\uparrow \downarrow}$,  (c) $\tilde \chi_{\uparrow \uparrow}$. Large dashed lines is $\tilde \chi_b$ in the three panels. Short dashed lines are the average pair radius (a) $\xi_{BCS}$ and the correlation lengths (b) $\xi_{\uparrow \downarrow}$ and (c) $\xi_{\uparrow \uparrow}$, as described in Section III.} \label{fig3}
\end{figure}

Figure \ref{fig3} shows the dimensionless $\tilde \chi_{BCS}$, $\tilde \chi_{\uparrow \downarrow}$,  $\tilde \chi_{\uparrow \uparrow}$  and $\tilde \chi_b$ as functions of $1/k_F a$, the first three ones plotted with solid lines and the latter with a large dashed line. We note first that in the BEC regime all four exponential decaying lengths behave essentially in the same way, indicating that the dominating length scales is that of $\chi_b$ (below we give their analytic asymptotic expression). On the other hand, in the BCS limit, while $\tilde \chi_{\uparrow \downarrow}$ and $\tilde \chi_{\uparrow \uparrow}$ appear very close to $\tilde \chi_b$, $\tilde \chi_{BCS}$ shows definitely a different behavior, also appearing divergent but at a much slower pace. As we will see below, when we analyze the mean pair size and correlation lengths, also shown in Figure \ref{fig3} with dashed lines, this departure makes a profound difference and gives rise to a richer picture of the structure of the mixture gas.

\begin{figure}[htbp]
\begin{center}
\includegraphics[width=0.8\linewidth]{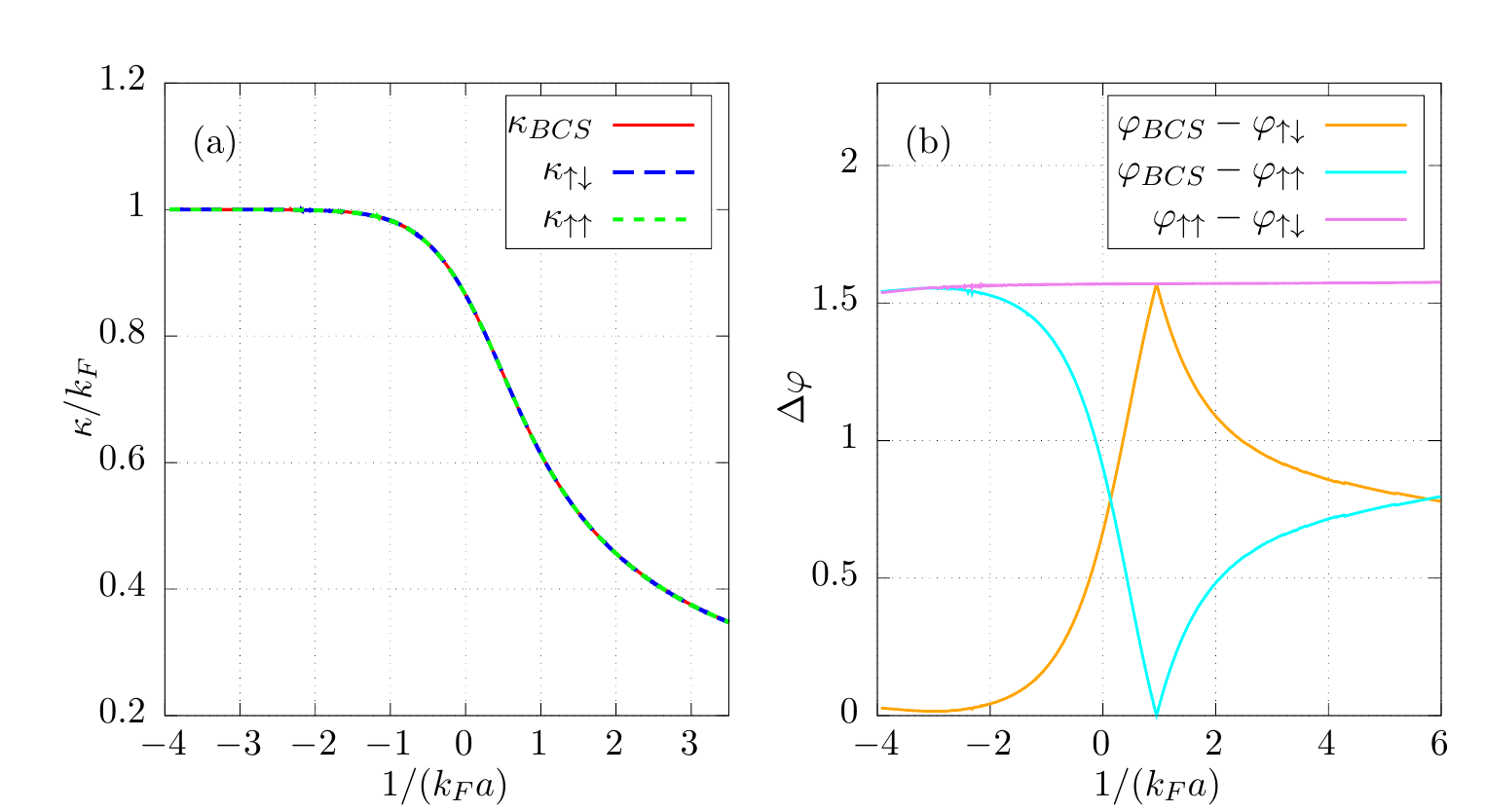}
\end{center}
\caption{(Color online) (a) Characteristic wavevector $\kappa$ and (b) phase differences in the long distance behavior of the respective pair wavefunction and density correlations, see Fig. \ref{fig2}.} \label{fig4}
\end{figure}

{\bf Characteristic wavenumbers and phases}. Two other important spatial features can be extracted from the pair probability density and the correlation functions: these are the characteristic wavenumber $\kappa$ of the oscillatory function and its phase $\varphi$. In Fig. \ref{fig4}(a), the plot of $\kappa_{BCS}$, $\kappa_{\uparrow \downarrow}$ and $\kappa_{\uparrow \uparrow}$ as functions of $1/k_Fa$, shows the striking conclusion that, within our numerical precision, they are exactly the same. This perhaps should not be surprising for the ${\uparrow \downarrow}$ and  ${\uparrow \uparrow}$ density correlations, since there must be an spatial correlation of the spin species due to pairing. In the BCS limit the characteristic wavenumber approaches the Fermi momentum $k_F$, an expected result \cite{LandauII,Giorgini-Stringari}, yet for the BEC limit, it appears to slowly vanish. The latter result appears to be in agreement with the fact that in such a limit, the ${\uparrow \downarrow}$ atoms form molecules and the gas should appear completely uncorrelated for large distances, as in any common gas. However, by looking at Figure \ref{fig4}(b), where the phase differences are shown, $\varphi_{\uparrow \uparrow}-\varphi_{\uparrow \downarrow}$ and $\varphi_{BCS} - \varphi_{\uparrow \downarrow}$, we observe that the former is always $\pi$ while the latter changes as a function $1/k_Fa$. This perfect phase difference can already be observed at short distances in the (a)-upper panel of Fig. \ref{fig2}. Thus, the correlations $G_{\uparrow \uparrow}(r)$ and $G_{\uparrow \downarrow}(r)$ show a very deep structure not only of the $\uparrow \downarrow$ pairs, but of the whole gas mixture, particularly in the BCS side: as $1/k_Fa \to -\infty$, the decaying lengths $\chi$ and the correlation lengths $\xi$ (see following Section) for both correlations diverge in the same way, indicating that those quantities become irrelevant in determining the structure. Hence, the equality of $\kappa_{\uparrow \downarrow}=\kappa_{\uparrow \uparrow}$ and the constancy of the phase difference $\varphi_{\uparrow \uparrow}-\varphi_{\uparrow \downarrow}$, indicate an average alternating shell structure from the perspective of any given atom. Although this structure remains in the BEC side, it is severely diminished by the vanishing of both $\chi_{\uparrow \downarrow}$ and $\chi_{\uparrow \uparrow}$.

\section{The mean pair radius and the correlation lengths}

Although all the spatial information we seek is in principle contained in the full distribution functions described in the previous section, there is a very important associated single quantity that serves to characterize their overall behavior. This is the second moment of the spatial distribution which, in the case of the BCS pair probability distribution function yields the mean pair radius, while in the correlation function gives their corresponding correlation lengths. These are defined as follows,
\begin{equation}
\xi^2 =\frac{ \left|\int r^2 \rho(\vec r) d^3r\right|}{\left|\int \rho(\vec r) d^3r\right|} ,\label{xi}
\end{equation} 
where $\rho(\vec r)$ is either the BCS pair probability density distribution $|\phi_{\rm BCS}(\vec r)|^2$, the pair correlation function $G_{\uparrow \downarrow}(\vec r)$, or the same spin correlation function $G_{\uparrow \uparrow}(\vec r)$. In the Appendix we provide details for the calculation of the different lenghts $\xi$. Similarly to the procedure in obtaining Eqs. (\ref{phi}), (\ref{eq_correlador_antiparalelo}) and (\ref{eq_correlador_paralelo}), one can also find closed analytical expressions for the corresponding lengths in terms of hypergeometric functions. In the case of the ${\uparrow \downarrow}$ and ${\uparrow \uparrow}$  correlation lengths, the final result can be written explicitly. Of these two, the former has already been reported in Refs. \cite{Pistolesi,Strinati-review}.

We quote here the analytical expressions for the average pair radius $\xi_{BCS}$ and the correlation lengths $\xi_{\uparrow \downarrow}$ and $\xi_{\uparrow \uparrow}$:
\begin{equation}
\xi_{\rm BCS}^2 = \frac{\hbar^2}{m (\mu^2 + \Delta^2)^{1/2}}\left\{\frac{1}{\sqrt{2}(1-z)(1+z)^{1/2}} \frac{1+2z-z^2-2z^3 + 2\sqrt{2}(1+z)^{1/2}\left( {\rm P}_{\frac{5}{2}}(z)- z {\rm P}_{\frac{3}{2}}(z)\right)}{- {\rm F}\left(\frac{7}{2},-\frac{5}{2};2,\frac{1-z}{2}\right) +z {\rm F}\left(\frac{5}{2},-\frac{3}{2};2,\frac{1-z}{2}\right)} \right\}.\nonumber\label{xiBCS}
\end{equation}
\begin{equation}
\xi_{\uparrow \downarrow}^2 = \frac{1}{8} \frac{\hbar^2}{2m} \frac{1}{\left(\mu^2 + \Delta^2\right)^{1/2}} \frac{5+8z+3z^2}{(1+z)^2} .
\end{equation}
\begin{equation}
\xi_{\uparrow \uparrow}^2 = \frac{3}{8} \frac{\hbar^2}{2m} \frac{1}{\left(\mu^2 + \Delta^2\right)^{1/2}} \left(\frac{1-z}{1+z}\right)
\end{equation}
In Fig. \ref{fig2} we plot $\tilde \xi_{BCS}$,  $\tilde \xi_{\uparrow \downarrow}$ and  $\tilde \xi_{\uparrow \uparrow}$, as functions of $1/k_Fa$ with short dashed lines in panels (a), (b) and (c) respectively. Their asymptotic limits, including that of the binding-energy pair distribution, in the BCS limit, $z \to -1$ and $1/k_Fa \to -\infty$, are
\begin{eqnarray}
\tilde \xi_{\rm BCS} &\approx& \frac{\sqrt{105}}{2} \left(\psi\left(\frac{7}{2}\right)-\psi\left(\frac{5}{2}\right) \right)^{1/2}= \sqrt{\frac{21}{2}}  \nonumber \\
\tilde \xi_{\uparrow \downarrow} & \approx & \frac{1}{\sqrt{2}}\frac{e^{2}}{8 }  e^{-\frac{\pi}{2 k_F a}} \nonumber \\
\tilde \xi_{\uparrow \uparrow} & \approx &\sqrt{\frac{{3}}{{2}}}\> \frac{e^{2}}{8 }  e^{-\frac{\pi}{2 k_F a}} \nonumber \\
\tilde \xi_{b} & \approx &\frac{2}{\sqrt{3}}\> \frac{e^{2}}{8 }  e^{-\frac{\pi}{2 k_F a}}
\label{xiBCSl}
\end{eqnarray}
where $\psi(x)$ is the digamma function. Note that in the BCS limit $\tilde \xi_{\rm BCS}$ reaches a finite limit. Thus, while $|\phi_{\rm BCS}(\vec r)|^2$ does become long ranged in the limit, we recall that $\tilde \chi_{\rm BCS}$ appears to bend in the BCS limit, see Fig. \ref{fig2} panel (a), growing slower than $\tilde \chi_{\uparrow \downarrow}$ and $\tilde \chi_{\uparrow \uparrow}$. As a consequence, the numerator and denominator of Eq. (\ref{xi}) both diverge as $\sim \tilde \Delta^{-1}$, yielding a finite $\tilde \xi_{BCS}$. Thus, from the perspective of the pair wavefunction, the pair size reaches a finite value in the BCS limit. This is in stark contrast to the ${\uparrow \downarrow}$ and ${\uparrow \uparrow}$ density correlation lengths which indeed diverge. However, there is no contradiction since the latter are correlation lengths and their divergence are in agreement of the expected notion that the correlation functions become long range. This indicates that while average pair radius is finite, the algebraic decay of the wavefunction allows for pairs of all sizes. We also note that, up to numerical factors of order one, $\tilde \xi_{\uparrow \downarrow} \approx \tilde \xi_{\uparrow \uparrow} \approx \tilde \xi_b \sim \tilde \Delta^{-1}$, also expressing the known fact that the gap is the relevant energy scale in the BCS limit, \cite{Ketterle-review} which agrees with the Pippard Coherence length \cite{Pistolesi}.

In the BEC limit, $z \to +1$ and $1/k_Fa \to + \infty$, we now have the asymptotic behavior,
\begin{eqnarray}
\tilde \xi_{\rm BCS} &\approx& \frac{1}{\sqrt{2}} \> k_F a \nonumber \\
\tilde \xi_{\uparrow \downarrow} & \approx & \frac{1}{\sqrt{2}}\>k_F a \nonumber \\
\tilde \xi_{\uparrow \uparrow} & \approx & \sqrt{\frac{1}{2\pi}}\> \left(k_Fa\right)^{5/2} \nonumber \\
\tilde \xi_{b} & \approx & \frac{1}{\sqrt{2}}\> k_Fa .\label{xiBEC}
\end{eqnarray}
While now $\tilde \xi_{BEC} \approx \tilde \xi_{\uparrow \downarrow} \approx \tilde \xi_b \sim |\tilde \mu|^{-1/2}$, showing that the relevant energy is $\tilde \mu$ and no longer the gap, the different behavior is shown by $\tilde \xi_{\uparrow \uparrow}$ which vanishes much faster than the others, as can be seen in Fig. \ref{fig2} panel (c). This can be interpreted as the lost of fermionic correlation between ${\uparrow \uparrow}$ and ${\downarrow \downarrow}$ atoms, since they now are part of bosonic indistinguishable molecules. This is in line with the attained BEC character of the gas in such a limit. To reinforce this point, the decorrelation observed in the BEC side is accompanied by the vanishing of the characteristic wavevectors $\kappa_{\uparrow \downarrow}$ and $\kappa_{\uparrow \uparrow}$, namely, the wavelength of the nested structure gets diluted as the pairs become tighter.

We point out that $g_{\uparrow \downarrow}(\vec r)$, the Fourier transform of $u_{\vec k}v_{\vec k}$ and whose square yields the pair correlation function $G_{\uparrow \downarrow}(\vec r)$, has also been identified as the ``pair wavefunction'' because it obeys a Schr\"odinger-like equation, as can be obtained from the gap equation, Eq.(\ref{gap}) \cite{Ketterle-review,Strinati-review,Leggett}. In this regard, it is considered that $g_{\uparrow \downarrow}(\vec r)$ and its associated correlation length $\xi_{\uparrow \downarrow}$ bears the Cooper pair structure. Indeed, Strinati and collaborators \cite{Strinati-review,Pistolesi,Marini} have thoroughly studied this function and its length not only in the zero-temperature limit as here but also at finite temperatures. It is thus of interest to point out that the correlation function $G_{\uparrow \downarrow}(\vec r)$, apart from its spatial oscillations, it is the one that most closely can be related to the binding-energy wavefunction along the whole crossover, succinctly summarizing the physical overall description of pairs. In this light, we believe that our work adds to the understanding of the structure of the mixture already gained by those studies.

\section{Final Remarks}

In summary, we have addressed the spatial structure of the gas mixture of two different fermionic species through the BEC-BCS crossover at zero-temperature mean-field level. We analyze the BCS pair wavefunction that enters the many body BCS ansatz and the equal ${\uparrow \uparrow}$ and pair ${\uparrow \downarrow}$ density correlation functions. 
By accurately calculating those functions we find them for all values of the corresponding distance pairs. For large distances we fit their exponential decay length, oscillatory behavior and relative phases. Moreover, we find exact analytical expressions for the second-moment lengths of the corresponding functions. These three distributions provide, in a complementary way, a quite complete overall description regarding the spatial structure of the mixture that, in principle, can be experimentally tested \cite{Ketterle-review,schunckdetermination}.

There are, however, natural extensions of this study that should be addressed. One is the consideration of realistic interatomic finite-range potentials \cite{Caballero,Neri} and the other is the inclusion of finite temperatures \cite{Palestini} still at the level of a mean-field description. For the case of finite temperatures there are already solid advances specially for the ${\uparrow \downarrow}$ correlation function \cite{Strinati-review,Palestini}. In this context there is an additional length, the {\it phase} length, which is associated to the gaussian fluctuations of the order parameter \cite{Marini}. Although such an analysis is beyond the scope of the present work, it is of interest to point out that such a quantity diverges in both limits, as $\tilde \Delta^{-1}$ in the BCS limit and as $|\tilde \mu|^{1/2}$ in the BEC one, taking its smallest value near unitarity, in contrast to the lengths here discussed. This suggests, as a natural continuation of the present work, to look at the fluctuations of all distributions here studied, even at the gaussian level.

Our calculations can be contrasted with the ones used in Monte Carlo methods. The structure of the correlators that we found may contribute to give physical insight for new variational functions \cite{Pessoa-contact}. Also comparing with those Monte Carlo calculations it can be seen that the correlation lengths obtained here are a lower bound of the correlation lengths of the complete many body Hamiltonian \cite{Giorgini-Stringari}.

To conclude we would like to briefly address the consequences of the long range behavior near unitarity and in the BCS limit on current experimental studies with confined ultracold gases. Typically, specially in the BEC side \cite{Hulet,Zhou,Nascimbene} and in gases with bosons such as $^{87}$Rb \cite{Bagnato} and $^{7}$Li, \cite{Navon}, the local density approximation has been shown to be quite accurate. This can be understood as a result that the density correlation length is (much) smaller than the size of the system. However, this may not be true in the near and deep BCS regions. That is, those vapor clouds may still be of relative mesoscopic size and, therefore, the usual thermodynamic description could not be directly applied to them. This deserves a further and careful analysis of the different length scales involved in those experiments.

\section*{Appendix: Analytical expressions and contour deformation}

\subsection*{Analytical expressions for thermodynamic functions and correlation lengths}

In the expressions (\ref{eq_ecuacion_gap_k}), (\ref{eq_ecuacion_num_k}) and (\ref{eq_ecuacion_ene_k}), the difficulty of the $\vec k$-integrals lie in the handling of the factor $\left((\epsilon_k-\mu)^2 + \Delta^2\right)^\beta$, when $\beta = -3/2, -1/2, 1/2, \dots$, a positive or negative semi-integer. The suggestion is to make the following change of variables  and rearrangements \cite{trascendental}
\begin{eqnarray}
\left((\epsilon_k-\mu)^2 + \Delta^2\right)^\beta &=& (\mu^2 + \Delta^2)^\beta \left(x^2 + 2xz +1\right)^\beta \nonumber \\ & = & (\mu^2 + \Delta^2)^\beta \left(1 + x\right)^{2\beta} \left( 1 - \frac{2 (1-z) x}{(1+x)^2}\right)^\beta \label{truco}
\end{eqnarray}
where $x = \epsilon_k/\sqrt{\mu^2 + \Delta^2}$ and $z = - \mu/\sqrt{\mu^2 + \Delta^2}$ and integrate over $x$. Since $x/(1+x)^2 < 1$ for $x \to 0$ and $x \to \infty$, one can perform a series expansion in powers of $2(1-z) x/(1+x)^2$ of expression (\ref{truco}). Then, all the resulting integrals can be rearranged as factors of a power series in $(1-z)/2$ that can be integrated term by term. This can a be a lengthy exercise but it yields a series of convergent integrals, which are all Beta functions. The resulting series can be cast in terms of hypergeometric functions,
\begin{equation}
{\rm F}(a,b;c,\frac{1-z}{2}) = \sum_{n=0}^{\infty} \frac{(a)_n (b)_n}{n! (c)_n} \left(\frac{1-z}{2}\right)^n ,
\end{equation}
where $(d)_n$ are Pochhammer symbols. Depending on $a$, $b$ and $c$, in some cases the hypergeometric functions  can be written in terms of Legendre functions and in others can be written explicitly \cite{trascendental,mathematica}.

For the calculation of the lengths $\xi$, Eq.(\ref{xi}), we first write the distribution $\rho(\vec r)$ in terms of its Fourier expression. For instance, for $\rho(\vec r) = |\phi_{BCS}(\vec r)|^2$, the length $\xi$ can be recast as,
\begin{eqnarray}
\xi_{\rm BCS}^2 & = & \frac{\int r^2 |\phi_{\rm BCS}(\vec r)|^2 d^3r}{\int |\phi_{BCS}(\vec r)|^2 d^3r} \nonumber \\
& = & \frac{\int |\nabla_k \tilde \phi_{\rm BCS}(\vec k)|^2 d^3 k}{\int |\tilde \phi_{BCS}(\vec k)|^2 d^3k}
\end{eqnarray}
where $\tilde \phi_{\rm BCS}(\vec k) = v_{\vec k}/u_{\vec k}$ is the Fourier transform of $\phi_{\rm BCS}(\vec r)$. One integrates the expression in the second line using the procedure described in the previous paragraph. Similarly for $\xi_{\uparrow \downarrow}$ and $\xi_{\uparrow \uparrow}$.

\subsection*{Contour deformation for the calculation of integral Eq.(\ref{fr})}

Here we show the steps to make the contour integration around the branch cuts in Eq. (\ref{fr}). In the following we will use the pair wave function $v_{\vec{k}}/u_{\vec{k}}$, but the same procedure can be used for $u_{\vec{k}}v_{\vec{k}}$ and $v_{\vec{k}}^2$.\\

\begin{figure}
\includegraphics[scale=0.9]{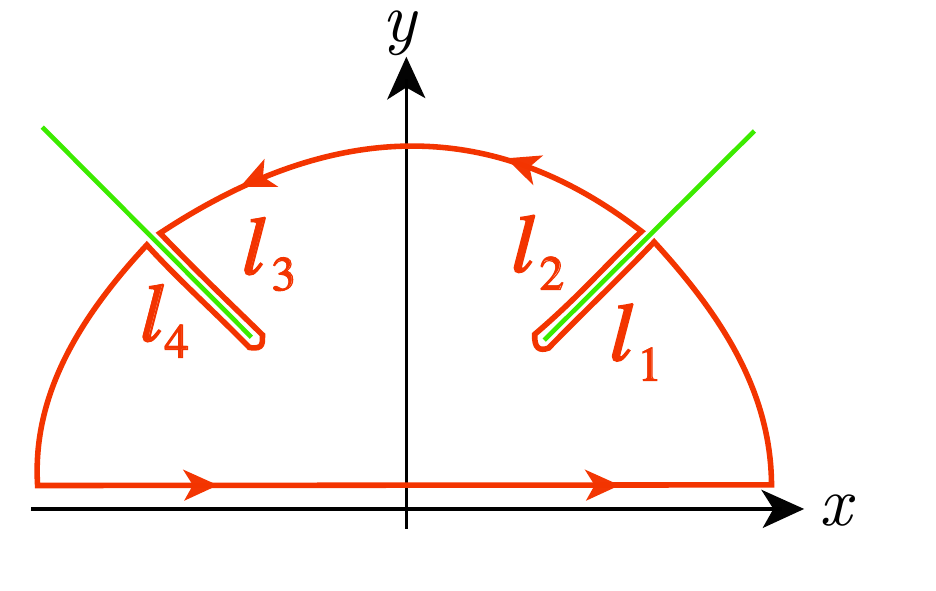}
\caption{\label{fig5} Illustration of the contour used in the Cauchy integral formula, with $\kappa = x+ i y$. Surrounding the branch cuts give four paths denoted by $l_i$, with $i=1,2,3,4$.}
\end{figure}

Using the gap $\Delta = \hbar^2 k_{\Delta}^2/2m$ to adimensionalize and defining $\mu_{\Delta}= \mu / \Delta$, the pair wave function is given by the following equation

\begin{equation}
(k_\Delta r) \phi(k_\Delta r) = \frac{k_\Delta^3}{4\pi^2i}\>I(k_\Delta r),
\end{equation}
where we have arranged the equation to focus only in the integral
\begin{equation}\label{eq_ecuacion_estudiar_apendice}
I(k_\Delta r) = \int_{- \infty}^{\infty}  \kappa [\sqrt{(\kappa^2 - \mu_\Delta)^2 + 1} - (\kappa^2 - \mu_\Delta)] e^{i \kappa (k_\Delta r)} \; d \kappa.
\end{equation}
The integrand has four branch cuts due to the square root. To calculate them we will use the principal branch of the complex logarithm. Then the points $\kappa$ in the complex plane belonging to the branch cuts satisfy the following equations
\begin{equation}\label{eq_condicion_branch_cut}
{\rm Re}[(\kappa^2 - \mu_\Delta)^2 + 1] \leq 0, \>\>\>\>{\rm and}\>\>\>\>
{\rm Im[}(\kappa^2 - \mu_\Delta)^2 + 1] = 0.
\end{equation}
Using $\kappa = x+iy$ it can be shown that the branch cuts correspond to points in the hyperbola $x^2 - y^2 = \mu_{\Delta}$ with their magnitud satisfying $|\kappa|^2 \geq \sqrt{\mu_{\Delta}^2 + 1}$. 
Then, choosing a contour ${\cal C}$ which surrounds the branch cuts in the upper half plane like the one shown in Fig. \ref{fig5}, the Cauchy theorem allows us to conclude that $\oint_{\cal C} f(\kappa) d\kappa = 0$, where
\begin{equation}
f(\kappa) =  \kappa \sqrt{(\kappa^2 - \mu_\Delta)^2 + 1} \> e^{i \kappa (k_\Delta r)}.
\end{equation}
The term $\kappa (\kappa^2-\mu_{\Delta})e^{i \kappa (k_\Delta r)}$ can be removed from the integrand, as it is analytic. Hence, it can be concluded that
\begin{equation}
\begin{split}
\int_{-\infty}^\infty f(\kappa) d\kappa=  \,2 \int_{l_2} \kappa \; i \; \sqrt{|(\kappa^2 - \mu_\Delta)^2 + 1 |\;}\,  e^{i \kappa (k_\Delta r)} \; d\kappa+ \; 2 \int_{l_4} \kappa  \; i \sqrt{|(\kappa^2 - \mu_\Delta)^2 + 1 |\;} \, e^{i \kappa (k_\Delta r)} d\kappa.
\end{split}\label{intk}
\end{equation}
with the branch cuts parametrized by $\gamma_2(t) = t+i\sqrt{t^2 - \mu_\Delta}$ and 
$\gamma_4(t) = -t+i\sqrt{t^2 - \mu_\Delta}$, with $t \in [t_0, \infty)$ and 
\begin{equation}
t_0 = \Bigg( \dfrac{\mu_\Delta + (\mu_\Delta^2+1)^{1/2}}{2} \Bigg)^{1/2}.
\end{equation}

Substitution of the parametric equations into Eq. (\ref{intk}) and identifying the integral $I(k_{\Delta}r)$ with the integral of $f(\kappa)$ in the reals we arrive at the desired result,
\begin{equation}
I(k_{\Delta}r)  = -4 i \int_{t_0}^\infty \Bigg( \frac{2t^2 - \mu_\Delta}{\sqrt{t^2 - \mu_\Delta}} \Bigg) \sqrt{  4t^2 (t^2 - \mu_\Delta) -1\;} \; \exp \Big[   - \sqrt{t^2 - \mu_\Delta}  (k_\Delta r )\Big] \; \sin (t (k_\Delta r ) ) \; dt.
\end{equation}

\begin{acknowledgments}

We thank support from grants CONACYT  255573 and UNAM PAPIIT-IN105217. JCOJ acknowledges support from a CONACYT scholarship.

\end{acknowledgments}

\bibliography{bibliography-new}

\end{document}